\documentclass[11pt,showkeys,superscriptaddress,aps,prd,preprint]{revtex4-1}
\usepackage{amsmath}
\usepackage{amssymb}
\usepackage{graphicx}

\newcommand{\bea}{\begin{eqnarray}}
\newcommand{\eea}{\end{eqnarray}}

\begin{document}

\title{Galilean covariance, Casimir effect and Stefan-Boltzmann law at finite temperature}
\author{S. C. Ulhoa }
\email{sc.ulhoa@gmail.com} \affiliation{Instituto de F\'{i}sica,
Universidade de Bras\'{i}lia, 70910-900, Bras\'{i}lia, DF,
Brazil.}

\author{A. F. Santos}
\email{alesandroferreira@fisica.ufmt.br}
\affiliation{Instituto de F\'{\i}sica, Universidade Federal de Mato Grosso,\\
78060-900, Cuiab\'{a}, Mato Grosso, Brazil.}
\affiliation{Department of Physics and Astronomy, University of Victoria,
3800 Finnerty Road Victoria, BC, Canada.}

\author{Faqir C. Khanna\footnote{Professor Emeritus - Physics Department, Theoretical Physics Institute, University of Alberta - Canada}}
\email{khannaf@uvic.ca}
\affiliation{Department of Physics and Astronomy, University of Victoria,
3800 Finnerty Road Victoria, BC, Canada.}

\date{\today}

\begin{abstract}

The Galilean covariance, formulated in 5-dimensions space, describes the non-relativistic physics in a way similar to quantum field theory.  Using a non-relativistic approach the Stefan-Boltzmann law and the Casimir effect at finite temperature for a particle with spin zero and $1/2$ are calculated. The Thermo Field Dynamics is used to calculate the temperature effect.

\end{abstract}



\maketitle
\section{Introduction}

The non-relativistic physics is described by Galilean covariance. There are numerous  applications, for instance in condensed matter physics \cite{Morawetz:2013tka, Moroz:2015jla}, in superfluidity phenomenon \cite{Schmitt:2014eka} among others. The algebraic structure underlying Galilean covariance is more intricate than its relativistic counterpart. Still it is possible to write a geometric formulation of Galilean transformations.

The Galilean covariant formalism consists in extending the ordinary space-time by adding an extra dimension. This involves its embedding in a five-dimensional, (4 + 1), de Sitter space. Mathematically, this corresponds to a central extension of the Galilean group, and thereby enables us to deal only with vector representations. Thus the Galilean vector is defined as $x^\mu=({\bf x}, t, s )$, where $s$ represents an extra dimension. The momentum vector $p^\mu$, in 5-dimensions, is interpreted physically as: $\vec{p}$, $E$ and $m$.

The Galilean symmetry is used to understand the electric and magnetic limits of electromagnetism \cite{LeBellac1973}. An extension to describe non-relativistic particles with any spin has been developed \cite{Lévy-Leblond1967}. A guide to the tensor structure of the Galilean transformations involving the extra dimension was introduced by Takahashi \cite{Takahashi:1988jz, PROP:PROP2190360106, PROP:PROP2190371203}. Since then numerous applications have been considered: tensor analysis has been developed \cite{Santana01031998}, non-relativistic Lagrangian field models have been constructed \cite{0305-4470-37-41-011}, quantization of non-relativistic fields has been analyzed \cite{Abreu2003244, deMontigny:2007fqc}, a covariant Galilean gravity theory has been written \cite{Ulhoa:2009at}, weak-field approximation of locally Galilean invariant gravitational theories has been examined \cite{Cuzinatto:2009aq}, among others.

The intent of this article is to obtain the non-relativistic counterpart of the Stefan-Boltzmann law and Casimir effect for a free particle of mass $m$ obeying the Schroedinger and Pauli-Schroedinger equations. The Galilean covariance approach provides a well defined way to understand phenomenon at low energies following the insights of quantum field theory.

There are three ways to understand role of temperature on Stefan-Boltzmann law and Casimir effect. First the Matsubara formalism is applied to system in equilibrium by associating temperature with a complex time coordinate \cite{Matsubara01101955}. Secondly the closed time path formalism \cite{Schwinger:1960qe, Moller:1960cva} is the real time formalism which leads to a doubling of degrees of freedom of systems. It can be applied to both equilibrium and non-equilibrium phenomena. Finally there is the Thermo Field Dynamics (TFD)~\cite{Takahashi:1996zn, 1982tfdc.book.....U, Umezawa:1993yq, Khanna:2009zz, Santana2000405, SantanaThermo} that doubles the Fock space and uses the Bogoliubov transformation to introduce temperature. This formalism is used for systems in equilibrium. In the present work this formalism will be used.

In section II, the Galilean covariance is briefly described. In section III, the Thermo Field Dynamics approach is introduced. In section IV, the non-relativistic analog of the Stefan-Boltzmann law and the Casimir effect for a free particle with spin zero and $1/2$ are presented. Finally in the last section some observations are given.

\section{Galilean Covariance} \label{gc}

Galilean Physics is present in all aspects of Classical Mechanics as well as in the quantum world since the Schroedinger equation respects the Galilei group. The Galilei transformations are given by
\begin{eqnarray}
\overline{\mathbf{x}} &=&R\mathbf{x+v}t+\mathbf{a,}  \nonumber \\
\overline{t} &=&t+b.  \label{t gal2}
\end{eqnarray}%
They define a group whose subgroups describe temporal, spatial translations, rotations and boost transformations. 

In order to construct a particle theory an unitary representation of the Galilean group is used. However it is not possible to achieve such a goal with above transformations (\ref{t gal2}). It is necessary to introduce the Galilean covariance \cite{Lévy-Leblond1967}.

In order to obtain covariant Galilean transformations let us recall the dispersal relation
\begin{equation}
p_{\mu }p^{\mu }=\mathbf{p}^{2}-2mE=0,  \label{disp5}
\end{equation}
with $p^{\mu}=(\mathbf{p},p^{4}=E,p^{5}=m)$, $p_{\mu }=g_{\mu\nu}p^{\nu }$ and
\begin{equation}
g_{\mu\nu}= \left(
  \begin{array}{ccccc}
    1 & 0 & 0 & 0 & 0 \\
    0 & 1 & 0 & 0 & 0 \\
    0 & 0 & 1 & 0 & 0 \\
    0 & 0 & 0 & 0 & -1 \\
    0 & 0 & 0 & -1 & 0 \\
  \end{array}
\right) \,. \label{1}
\end{equation}
Here $g_{\mu\nu}$ is the Galilean metric. There is a manifold endowed with such a metric whose coordinates transform as

\begin{equation}
x^{\mu}\,'=\Lambda^{\mu}\,_{\nu}  x^{\nu}+a^\mu
\,,\label{3.2}
\end{equation}
where $a^\mu=({\bf a}, b,0)$ and $\Lambda^{\mu}\,_\nu $ is given by

\begin{equation}
\Lambda^{\mu}\,_\nu= \left(
           \begin{array}{ccc}
             R & 0 & -\textbf{V} \\
             -\textbf{V}\cdot R & 1 & \frac{1}{2}V^2 \\
             0 & 0 & 1 \\
           \end{array}
         \right)\,. \label{3.3}
\end{equation}
The invariant line element is
\begin{equation}
dl^2=g_{\mu\nu}dx^{\mu}dx^{\nu}\,,\label{3.6}
\end{equation}
with $g_{\mu\nu}=\Lambda^{\alpha}\,_\mu \Lambda_{\alpha\nu}$. The fifth coordinate $x^5=s$ is canonically conjugate to mass. The Galilean physics lies on a null geodesic in a 5-dimensional space. Then $dr^2-2dtds=0$, which leads to $ds=\frac{{\bf v}}{2}\cdot {\bf dr}$. In this sense the fifth coordinate is associated to the other four. Hence the Galilean transformations in equations (\ref{t gal2}) are viewed as an embedding of the 4-dimensional physical space into the 5-dimensional space. The algebra of the Galilean group is

\begin{equation*}
\begin{array}{rcl}
\left[ M_{\mu \nu },M_{\rho \sigma }\right]  & = & i\left( \eta _{\mu \rho
}M_{\nu \sigma }+\eta _{\nu \sigma }M_{\mu \rho }-\eta _{\nu \rho }M_{\mu
\sigma }-\eta _{\mu \sigma }M_{\nu \rho }\right)  \\
{\left[ P_{\mu },M_{\rho \sigma }\right] } & = & -i\left( \eta _{\mu \rho
}P_{\sigma }-\eta _{\mu \sigma }P_{\rho }\right)  \\
{\left[ P_{\mu },P_{\nu }\right] } & = & 0.%
\end{array}%
\end{equation*}%
Here rotation, $M_{ij}\rightarrow \epsilon _{ijk}J_{k}$, translation, $P_{i}$, temporal translation, $P_{4}\rightarrow -H$, and boost, $M_{5i}=-M_{i5}\rightarrow K_{i}$, are defined.

The Schrodinger equation which has Galilean symmetry is written covariantly in the 5-dimensional space,
\begin{eqnarray}
&&i\frac{\partial}{\partial t}\Psi\left(  \mathbf{x},t\right) =-\frac{1}%
{2m}\nabla^{2}\Psi\left(  \mathbf{x},t\right) \nonumber\\
&&p^\mu p_\mu \Psi=0\,. \label{schro}%
\end{eqnarray}
This implies that a massless scalar field defined in the 5-dimensional Galilean manifold leads to the Schroedinger equation. A massless Dirac field leads to the Pauli-Schroedinger equation,
\bea
\gamma^\mu\partial_\mu\Psi=0.
\eea

\section{Thermo Field Dynamics}\label{TFD}

Thermo Field Dynamics (TFD) is introduced in a covariant formalism. The main idea in TFD formalism is a doubling of the Fock space ${\cal S}$ leading to ${\cal S}_T={\cal S}\otimes \tilde{\cal S}$, where $\tilde{\cal S}$ is the fictitious Fock space. For an arbitrary operator ${\cal X}$ the standard doublet notation is
\bea
{\cal X}^a=\left( \begin{array}{cc} {\cal X}\\
\xi\tilde{{\cal X}}^\dagger \end{array} \right),
\eea
with $\xi = -1$ for bosons and $\xi = +1$ for fermions. The map between the tilde $\tilde{\cal X}_i$ and non-tilde ${\cal X}_i$ operators is defined by the following tilde conjugation rules:
\bea
({\cal X}_i{\cal X}_j)^\thicksim = \tilde{{\cal X}_i}\tilde{{\cal X}_j}, \quad\quad (c{\cal X}_i+{\cal X}_j)^\thicksim = c^*\tilde{{\cal X}_i}+\tilde{{\cal X}_j}, \quad\quad ({\cal X}_i^\dagger)^\thicksim = \tilde{{\cal X}_i}^\dagger, \quad\quad (\tilde{{\cal X}_i})^\thicksim = -\xi {\cal X}_i\,.
\eea

The Bogoliubov transformation introduces thermal effects through a rotation between tilde and non-tilde variables. Using the doublet notation we get
\bea
\left( \begin{array}{cc} {\cal X}(\alpha)  \\ \tilde {\cal X}^\dagger(\alpha) \end{array} \right)={\cal B}(\alpha)\left( \begin{array}{cc} {\cal X}(k)  \\ \tilde {\cal X}^\dagger(k) \end{array} \right),
\eea
where the Bogoliubov transformation, ${\cal B}(\alpha)$, is given as
\bea
{\cal B}(\alpha)=\left( \begin{array}{cc} u(\alpha) & -v(\alpha) \\
\xi v(\alpha) & u(\alpha) \end{array} \right),
\eea
with $u^2(\alpha)+\xi v^2(\alpha)=1$. The temperature, $T$, is associated with the parameter, $\alpha$, as $\alpha=1/k_B T$, where $k_B$ is the Boltzmann constant.

As an example, consider the propagator for the scalar field given as
\bea
G_0^{(ab)}(x-x';\alpha)=i\langle 0,\tilde{0}| \tau[\phi^a(x;\alpha)\phi^b(x';\alpha)]| 0,\tilde{0}\rangle,
\eea
where,
\bea
\phi(x;\alpha)&=&{\cal B}(\alpha)\phi(x){\cal B}^{-1}(\alpha).
\eea
Here $a, b=1,2$ and $\tau$ is the time ordering operator. Using the thermal vacuum $|0(\alpha)\rangle={\cal B}(\alpha)|0,\tilde{0}\rangle$ the propagator becomes
\bea
G_0^{(ab)}(x-x';\alpha)&=&i\langle 0(\alpha)| \tau[\phi^a(x)\phi^b(x')]| 0(\alpha)\rangle,\nonumber\\
&=&i\int \frac{d^4k}{(2\pi)^4}e^{-ik(x-x')}G_0^{(ab)}(k;\alpha),
\eea
where
\bea
G_0^{(ab)}(k;\alpha)={\cal B}^{-1}(\alpha)G_0^{(ab)}(k){\cal B}(\alpha),
\eea
with
\bea
G_0^{(ab)}(k)=\left( \begin{array}{cc} G_0(k) & 0 \\
0 & \xi G^*_0(k) \end{array} \right),
\eea
and
\bea
G_0(k)=\frac{1}{k^2-M^2+i\epsilon},
\eea
where $M$ is the scalar field mass.

As physical quantities are given by the non-tilde variables we obtain
\bea
G_0^{(11)}(k;\alpha)=G_0(k)+\xi v^2(k;\alpha)[G^*_0(k)-G_0(k)],
\eea
where the generalized Bogoliubov transformation \cite{Khanna:2011gf} is given as
\bea
v^2(k;\alpha)=\sum_{s=1}^d\sum_{\lbrace\sigma_s\rbrace}2^{s-1}\sum_{l_{\sigma_1},...,l_{\sigma_s}=1}^\infty(-\eta)^{s+\sum_{r=1}^sl_{\sigma_r}}\,\exp\left[{-\sum_{j=1}^s\alpha_{\sigma_j} l_{\sigma_j} k^{\sigma_j}}\right],\label{BT}
\eea
with $d$ being the number of compactified dimensions, $\eta=1(-1)$ for fermions (bosons) and $\lbrace\sigma_s\rbrace$ denotes the set of all combinations with $s$ elements.

\section{Non-relativistic Stefan-Boltzmann law and Casimir effect}\label{results}

In this section the Stefan-Boltzmann law and Casimir effect at finite temperature are obtained for Schroedinger and Pauli-Schroedinger equations.

\subsection{Schroedinger equation}

In the framework of 5-dimension Galilean space the Schroedinger equation is represented by a massless, $M=0$, scalar field. Its Lagrangian density is given as
\bea
{\cal L}=\frac{1}{2}\partial_\mu\phi\partial^\mu\phi\,.
\eea
In order to calculate the Stefan-Boltzmann law and Casimir effect the energy-momentum tensor is 
\bea
T^{\mu\nu}=\partial^\mu\phi\partial^\nu\phi-g^{\mu\nu}{\cal L}\,.
\eea
To avoid divergences, the energy-momentum tensor is rewritten at different space-time points 
\bea
T^{\mu\nu}(x)=\lim_{x^\mu\rightarrow x'^\mu}\left(\partial^\mu\partial'^\nu-\frac{1}{2}g^{\mu\nu}\partial_\alpha\partial^\alpha\right)\tau[\phi(x)\phi(x')]\,.
\eea
Averaging on a vacuum state leads to the Green function \cite{Abreu2003244},
\bea
G_0(x^\mu-x'{^\mu})&=&\left\langle 0\left|\tau[\phi(x)\phi(x')]\right|0\right\rangle,\nonumber\\
&=&\frac{1}{(2\pi)^3}\theta(t-t')e^{-im(s-s')}\int d^3p\,e^{i\left[{\bf p}\cdot ({\bf{x}}-{\bf{x'}})-p^2(t-t')/2m\right]},\nonumber\\
&=&\left(\frac{m}{2\pi}\right)^{3/2}\frac{\sqrt{i}}{(t-t')^{3/2}}\theta(t-t')\times\nonumber\\
&\times &\exp \left\{-im (s-s')+\frac{im\left[(x-x')^2+(y-y')^2+(z-z')^2\right]}{2(t-t')}\right\}\,,
\eea
where $m$ is the particle mass. Using such a function average of every component of the energy-momentum tensor diverges in the limit $x^\mu\rightarrow x'^\mu$ when a zero temperature state is used. 

Calculating the energy-momentum tensor at finite temperature leads to
\bea
\left\langle 0(\beta)\left|T^{44(11)}\right| 0(\beta)\right\rangle&=&\lim_{x^\mu\rightarrow x'^\mu}\sum_{j=1}^{\infty} \partial_5\partial'_5\, G_0^{(11)}\left(x^\mu- x'^\mu-i\beta j n_4\right),
\eea
where $G_0^{(11)}\left(x^\mu- x'^\mu-i\beta j n_4\right)$ is the Green function at finite temperature with $\alpha=(0,0,0,\beta,0)$ and $n_4=(0,0,0,1,0)$. Then
\bea
\left\langle 0(\beta)\left|T^{44(11)}\right| 0(\beta)\right\rangle&=&\left[\frac{m^{7/2}}{(2\pi\beta)^{3/2}}\right]\zeta\left(\frac{3}{2}\right),
\eea
where $\zeta(3/2)$ is the Riemann Zeta function. This is the non-relativistic Stefan-Boltzmann law. It is interesting to compare this result with usual relativistic approach to such a calculation that yields 
\bea
E=\frac{\pi^2}{30}T^4.
\eea

Analogously the non-relativistic Casimir energy is
\bea
\left\langle 0(\beta)\left|T^{33(11)}\right| 0(\beta)\right\rangle&=&-\frac{1}{2}\lim_{x^\mu\rightarrow x'^\mu}\sum_{j,\gamma=1}^{\infty}\left(\partial_1\partial'_1+\partial_2\partial'_2+\partial_3\partial'_3-2\partial_4\partial'_5\right)\times\nonumber\\
&\times & G_0^{(11)}\left(x^\mu- x'^\mu-i\beta j n_4-2d\gamma n_3\right)\,,\nonumber\\
&=& m\left(\frac{m}{2\pi}\right)^{3/2}\sum_{j,\gamma=1}^{\infty}\left[\frac{1}{(j\beta)^{5/2}}+\frac{4md^2\gamma^2}{(j\beta)^{7/2}}\right]\exp \left(-\frac{md^2\gamma^2}{j\beta}\right)\,,
\eea
where $\alpha=(0,0,i2d,\beta,0)$, $n_3=(0,0,1,0,0)$ and $d$ is the distance between two parallel plates. This is the non-relativistic Casimir energy for spin zero particles at finite temperature. It is interesting to compare this result with usual relativistic approach to such a calculation that yields 
\bea
E=-\frac{1}{2\pi^2}\sum_{j,\gamma}\frac{(2d\gamma)^2-3(\beta j)^2}{\left[(2d\gamma)^2+(\beta j)^2\right]^3}.
\eea 

\subsection{Pauli-Schroedinger equation}

The Pauli-Schroedinger equation is described by the Dirac-like Lagrangian density in the Galilean manifold as
\bea
\mathfrak{L}=\frac{1}{2}\left(\bar{\Psi}\gamma^\mu \partial_\mu\Psi-\partial_\mu\bar{\Psi}\gamma^\mu\Psi\right)\,.
\eea
Similar to the previous case, the energy-momentum tensor is calculated at different points of space-time, then
\bea
T^{\mu\nu}=\lim_{x^\mu\rightarrow x'^\mu} \gamma^\mu \partial^\nu \tau\left[\bar{\Psi}(x')\Psi(x)\right]\,.
\eea
Averaging it on a vaccum state and using $\gamma^\mu\gamma_\mu=5$ we get
\bea
\left\langle 0\left|T^{\mu\nu}\right|0\right\rangle=\lim_{x^\mu\rightarrow x'^\mu}-5\,\partial^\mu\partial^\nu  G_0(x^\mu-x'{^\mu})\,.
\eea
In this form the components of the averaged energy-momentum tensor diverge. In a thermal vacuum state the energy is 
\bea
\left\langle 0(\beta)\left|T^{44(11)}\right| 0(\beta)\right\rangle&=&-5\lim_{x^\mu\rightarrow x'^\mu}\sum_{j=1}^{\infty}(-1)^{j+1}\partial_5\partial_5\,  G_0^{(11)}(x^\mu-x'{^\mu}-i\beta jn_4)\,\nonumber\\
&=&5m^2\left(\frac{m}{2\pi \beta}\right)^{3/2}\left[1-\frac{\sqrt{2}}{2}\right]\zeta\left(\frac{3}{2}\right)\,.
\eea
This represents the Stefan-Boltzmann law for the non-relativistic spin-half particle. The non-relativistic Casimir energy for a spin-half particle at finite temperature is 
{\small
\bea
\left\langle 0(\beta)\left|T^{33(11)}\right| 0(\beta)\right\rangle&=&-5\lim_{x^\mu\rightarrow x'^\mu}\sum_{j,\gamma=1}^{\infty}(-1)^{j+\gamma}\partial_3\partial_3\,  G_0^{(11)}(x^\mu-x'{^\mu}-i\beta jn_4-2d\gamma n_3)\nonumber\\
&=&5m\left(\frac{m}{2\pi \beta}\right)^{3/2}\sum_{j,\gamma=1}^{\infty}\frac{(-1)^{j+\gamma}}{j^{3/2}}\left[\frac{4md^2\gamma^2-j\beta}{j^2\beta^2}\right]\exp \left(-\frac{2md^2\gamma^2}{j\beta}\right).
\eea
}

\section{Conclusion} \label{con}

The Galilean covariance is defined in a framework with 5-dimensions. Such a space is to be viewed as an Euclidean space embedding in a (4,1) de Sitter space. The finite temperature effects are introduced using the Thermo Field Dynamics. The non-relativistic Stefan-Boltzmann law and Casimir effect at finite temperature for a free particle are obtained using the Schroedinger and Pauli-Schroedinger equations in the Galilean covariant formalism. The non-relativistic Stefan-Boltzmann law represents just the energy of a free particle with spin zero or $1/2$, while the non-relativistic Casimir effect means just the pressure of such a single particle.

\section*{Acknowledgments}

This work by A. F. S. is supported by CNPq projects 476166/2013-6 and 201273/2015-2. S. C. U. thanks the Funda\c{c}\~ao de Apoio $\grave{a}$ Pesquisa do Distrito Federal - FAPDF for financial support. We thank Physics Department, University of Victoria for access to facilities.


%

\end{document}